\documentclass[apj]{emulateapj}

\newcommand{\apjvec}[1]{\mbox{\boldmath{$#1$}}}

\newcommand{\vx}{\apjvec{x}}
\newcommand{\vcr}{\apjvec{r}}
\newcommand{\vmu}{\apjvec{\mu}}

\begin{document}

\title{Cross-Identification of Stars with Unknown Proper Motions}
\journalinfo{Submitted to ApJ}

\author{
Gy\"ongyi Kerekes\altaffilmark{1},
Tam\'as Budav\'ari\altaffilmark{2,1},
Istv\'an Csabai\altaffilmark{1,3},
Andrew J. Connolly\altaffilmark{4},
Alexander S. Szalay\altaffilmark{2}
}

\iffalse
\affil{Dept.\ of Physics of Complex Systems, E\"otv\"os Lor\'and University, P\'azm\'any P.\ s\'et\'any 1/A, Budapest, 1117, Hungary}
\affil{Dept.\ of Physics and Astronomy, The Johns Hopkins University, 3400 North Charles Street, Baltimore, MD~21218, USA}
\else
\altaffiltext{1}{E\"otv\"os Lor\'and University, Department of Physics of Complex Systems, P\'azm\'any P.\ s\'et\'any 1/A, Budapest, 1117, Hungary}
\altaffiltext{2}{Department of Physics and Astronomy, The Johns Hopkins University, 3400 North Charles Street, Baltimore, MD~21218, USA}
\altaffiltext{3}{Institute for Advanced Study, Collegium Budapest, Szenth\'aroms\'ag u. 2, H-1014 Budapest, Hungary}
\altaffiltext{4}{Department of Astronomy, University of Washington, 3910 15th Avenue NE, Seattle, WA~98195, USA}
\fi

\email{kerekes@complex.elte.hu, budavari@jhu.edu}

\shortauthors{Kerekes, Budav\'ari et~al.}
\shorttitle{Cross-Identification of Stars}

\begin{abstract}
The cross-identification of sources in separate catalogs is one
of the most basic tasks in observational astronomy.
It is, however, surprisingly difficult and generally ill-defined.
Recently \citet{pxid} formulated the problem in the realm of
probability theory, and laid down the statistical foundations of
an extensible methodology.
In this paper, we apply their Bayesian approach to stars
with detectable proper motion,
and show how to associate their observations.
We study models on a sample of stars in the Sloan Digital Sky Survey,
which allow for an unknown proper motion per object, and 
demonstrate the
improvements over the analytic static model.
Our models and conclusions are directly applicable to  upcoming
surveys such as PanSTARRS, the Dark Energy Survey, Sky Mapper, and the LSST,
whose data sets will contain hundreds of millions of stars observed
multiple times over several years.
\end{abstract}


\section{Introduction} \label{sec:intro}

At the heart of many astronomical studies today is the basic step of
catalog merging; combining
measurements from different time intervals, wavelengths, and
potentially separate instruments and telescopes.
Scientific analyses exploit these multicolor cross-matches to
understand the temporal and photometric nature of the underlying objects. 
In doing so they rely implicitly
on the quality of the associations, thus the cross-identification of sources is
arguably one of the most important steps in measuring the
properties of celestial objects.

In general, cross-matching catalogs is a difficult problem that cannot
really be separated from the
scientific question at hand.
An example of this is apparent when we consider the case of stellar
observations. Stars that move between observations, due to their
proper motions, are difficult to merge into multicolor sources (even
within a single survey). Yet without the multicolor information it
might not be possible to classify the source as a star in the first place.
With a new generation of surveys that will take large quantities of
multicolor photometry covering the Galactic Plane and observed over a
period of several years (e.g.\ the Panoramic Survey Telescope \& Rapid
Response System, PanSTARRS, and the Large Synoptic Survey Telescope,
LSST) it is clear that addressing these issues is becoming a serious
concern.

In the recent work of \citet{pxid} a general probabilistic formalism was
introduced that is extendable to arbitrarily complex models.
The beauty of the approach of Bayesian hypothesis testing
is that it clearly separates
the contributions of different types of measurements, e.g., the position on
the sky or the colors of the sources, yet, naturally combines them into
a coherent method.
It is a generic framework that provides the prescription for the calculations
that can be refined with more and more sophisticated modeling.

In this paper, we go beyond the simple case of stationary objects, and study
the cross-identification of point sources that move on the sky. Most importantly
we focus on stars that can be significantly offset between
the epochs of observations.
Although we only have loose constraints on their proper motions in general,
this prior knowledge is enough to revise our static models,
and work out the Bayesian evidence of the matches.
In Section~\ref{sec:bf} we introduce a class of models that allow for
changes in the position over time.
Section~\ref{sec:prior} deals with the a priori constraints on the
proper motions of the stars and their empirical ensemble statistics.
In Section~\ref{sec:res} we show the improvements over the static
model on actual observations of stars, and
Section~\ref{sec:sum} concludes our study.
Throughout this paper, we adopt the convention to use the capital
$P$ symbol for probabilities and the lower case $p$ letter for probability densities.

\section{Proper Motion} \label{sec:bf}

Conceptually, modeling the position of moving sources is straightforward.
The description combines the motion and the uncertainty of
the astrometric measurements.
The first question to answer is where on the sky one should expect to see an
object of a certain proper motion, if it had been in some known position
at a given time.
Next, we calculate the evidence that given detections
are truly observations of the same object.

\subsection{Multi-epoch Models}

The positional accuracy is characterized by a probability density function
(hereafter PDF) on the celestial sphere. In a given model $M$,
this $p(\vx|\vcr,M)$ function tells us where to expect $\vx$ detections of an object that is at
its true location $\vcr$.
Throughout this paper, we use 3-dimensional unit vectors for the positions on the sky, e.g., the
aforementioned $\vx$ and $\vcr$ quantities.
Usually the PDF is a very sharp peak and is assumed to be a normal
distribution with some angular accuracy $\sigma$. The correct generalization to directional
measurements is the \citet{fisher53} distribution,
\begin{equation}
F(\vx|\vcr,w)
  = \frac{w\,\delta(|\vx|\!-\!1)}{4\pi \sinh w}\ \exp \left( {w\,\vcr{}\vx} \right)
\end{equation}
whose shape parameter $w$ is essentially
$1/\sigma^2$ in the limit of large concentration; see details in \citet{pxid}.

The added complication comes from the fact that some objects are not stationary.
If a given star is at location  $\vcr$ now and has $\vmu$ proper motion 
then $\Delta{}t$ time later, it would be at some other position $\vcr'$
\begin{equation}
\vcr' = \vcr'(\Delta{}t; \vcr, \vmu)
\end{equation}
that is offset by a small displacement along a great circle.
By substituting this position into our astrometric model, we create
a new one $M'$ with the added proper motion, $\vmu$, and time 
difference, $\Delta{}t$, parameters.
\begin{equation}
p(\vx|\Delta{}t,\vcr,\vmu,M') = F\left(\vx|\vcr'(\Delta{}t; \vcr,\vmu),w\right)
\end{equation}
Naturally, there is nothing specific in this about the chosen characterization
of the astrometry;
one can use any appropriate PDF in place of the Fisher distribution instead.

\subsection{The Bayes Factor}

At the heart of the probabilistic cross-identification is the Bayes factor
used for hypothesis testing.
The question we are asking is whether our data $D$, a set of detected sources in separate
catalogs with positions $\{\vx_i\}$, are truly from the same object.
For every catalog, we know its epoch and its astrometry characterized by a known $p_i$ PDF.
Let $H$ denote
the hypothesis that assumes that all measured positions are observations of the same object,
and let $K$ denote its complement, i.e., any one or more of the detections might belong to a separate
object.
By definition, the Bayes factor is the ratio of the likelihoods of the two hypotheses we wish
to compare,
\begin{eqnarray} \label{eq:bf}
B(H,K|D) & = & \frac{p(D|H)}{p(D|K)}
\end{eqnarray}
that are calculated as the integrals over their entire parameter spaces.

If we assume that there is a single object behind the observations,
we can integrate over its unknown proper motion and position to
calculate
\begin{equation} \label{int1}
p(D|H) = \int\!\!\!d\vcr\!\!\int\!\!\!d\vmu\,p(\vcr,\vmu|H) \prod_{i=1}^{n} p_i(\vx_i | \Delta{}t_i,\vcr,\vmu, H)
\end{equation}
where
the joint likelihood of $H$ given the data is written as the product of the independent components
and $p(\vcr,\vmu)$ is the prior on the parameters, which is the subject of the following section.
The actual calculation of this likelihood depends on the prior and might only
be accessible via numerical methods.

The complementary hypothesis is more complicated in the sense that the model
has a set of independent objects with $\{\vcr_i,\vmu_i\}$ parameters, however,
the result of the calculation turns out to be much simpler.
Here the integral separates into the product of
\begin{equation} \label{int2}
p(D|K)  = \prod_{i=1}^{n} \int\!\!\!d\vcr_i\!\!\int\!\!\!d\vmu_i\,p(\vcr_i,\vmu_i|K)\,p_i(\vx_i | \Delta{}t_i,\vcr_i,\vmu_i, K)
\end{equation}
For each integral, we can select a reference time such that $\Delta{}t_i=0$,
hence the effect of the proper motion drops out, and
we arrive at the same result as the stationary case discussed by \citet{pxid}.

\section{Prior Determination} \label{sec:prior}

The proper motion really only shows up in the numerator of the Bayes factor
for assessing the quality of the association. The model is well-defined
but the integration domain is set by the joint prior that is yet to be
determined.
In general, the prior $p(\vcr,\vmu|H)$ can be very complicated for its dependence on the properties of the star. Simply put, brighter sources are likely to be closer, and hence, have a larger proper motion. More complicated is the effect of the color that is (along with its magnitude) a proxy for placing the star in different stellar populations with different dynamics. In this paper, we will not discuss these effects that will be a topic of future work. We also note that the prior can be a function of the time difference, $\Delta t$, to account for cases when the star travels far between observations to a new location with different source density. However, we expect this to be a small effect because the typical speed of stars and the usual time differences between observations today yield small displacements on the sky.

Using the basic properties of conditional densities, we can
write it as the product
\begin{equation} \label{prior}
p(\vcr,\vmu|H) = p(\vcr|H)\,p(\vmu|\vcr,H)
\end{equation}
where the first term is the prior on the position, e.g.,
the all-sky prior written with Dirac's $\delta$ symbol as
\begin{equation}
p(\vcr|H) = \frac{1}{4\pi}\,\delta(|\vcr|-1)
\end{equation}
and the more complicated second term describes the possible proper
motions as a function of location and optionally other properties.
The simplest possible model, after the stationary case, is to assume a
uniform prior on $\vmu$ up to some $\mu_{\rm{max}}$ limit independent of
the location, i.e.,
\begin{equation} \label{eq:uconst}
p(\vmu|\vcr,H) =  \left\{\begin{array}{c l}
           1 \big/ \pi\mu_{\rm{max}}^{2}  & \quad \mbox{if\ $|\vmu|<\,\mu_{\rm{max}}$}\\
           0 & \quad \mbox{otherwise}\\ \end{array} \right.
\end{equation}
We will use this simple prior for comparison in addition to the stationary case, 
where $\vmu$ is assumed to be negligible. 
%

\begin{figure}
\epsscale{1.2}
\plotone{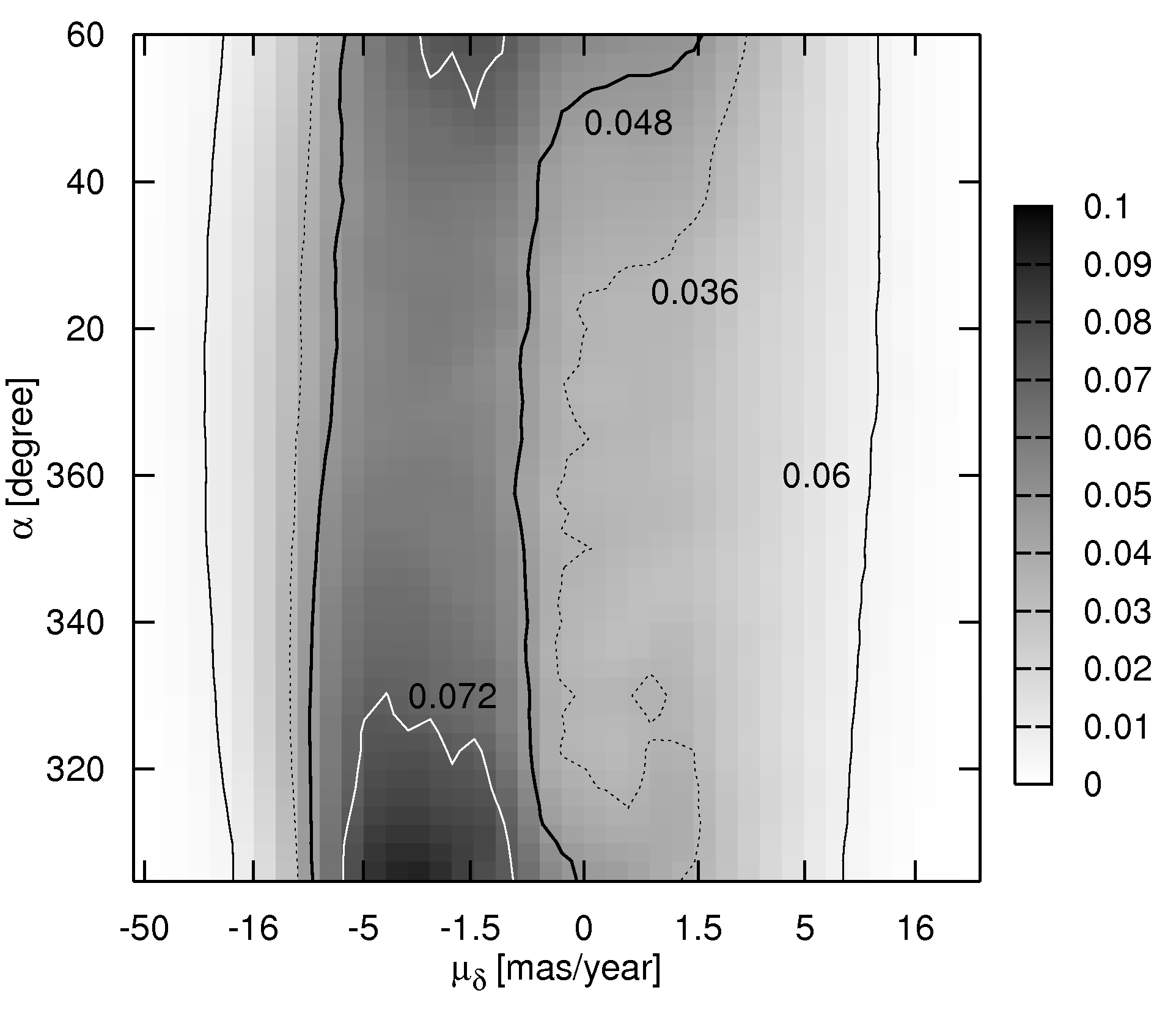}
\caption{Illustration of the first term of the prior (\ref{PM1}) in the 
footprint of SDSS Stripe 82. The $\mu_{\delta}$ axis has an asinh() scale. 
As one moves into the direction of Galactic Plane 
(upwards and downwards on the figure), the probability distribution gets 
narrower since the velocity dispersion of stars decreases. }
\label{prior1}
\end{figure}

\begin{figure*}
\epsscale{1.1}
\plotone{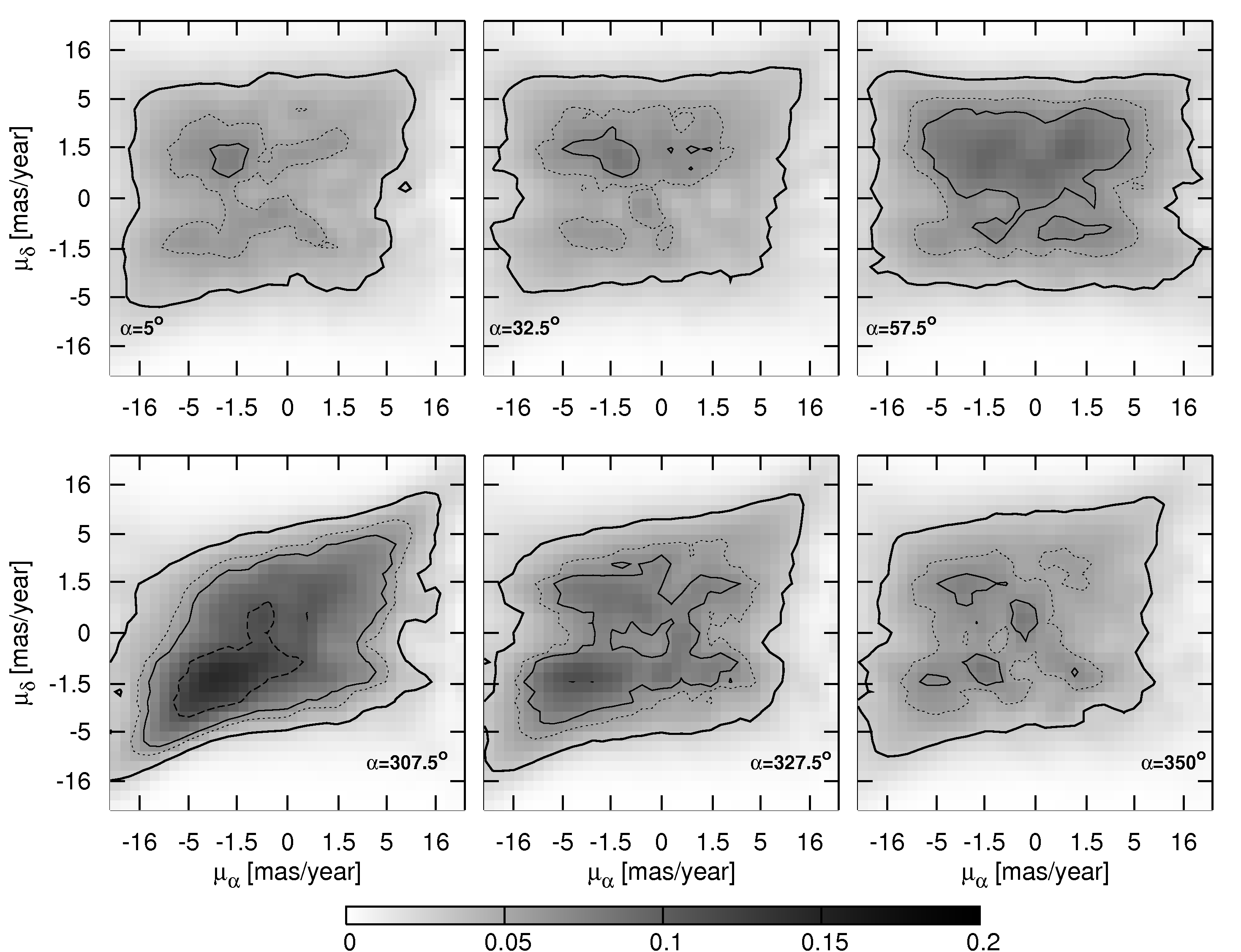}
\caption{Illustration of the second term of the prior (\ref{PM1}). 
The function is 4-dimensional; 
The 6 different panels represent slices of the 4-dimensional 
PDF in the R.A.\ direction.}
\label{prior2}
\end{figure*}

\begin{figure*}
\epsscale{1.1}
\plotone{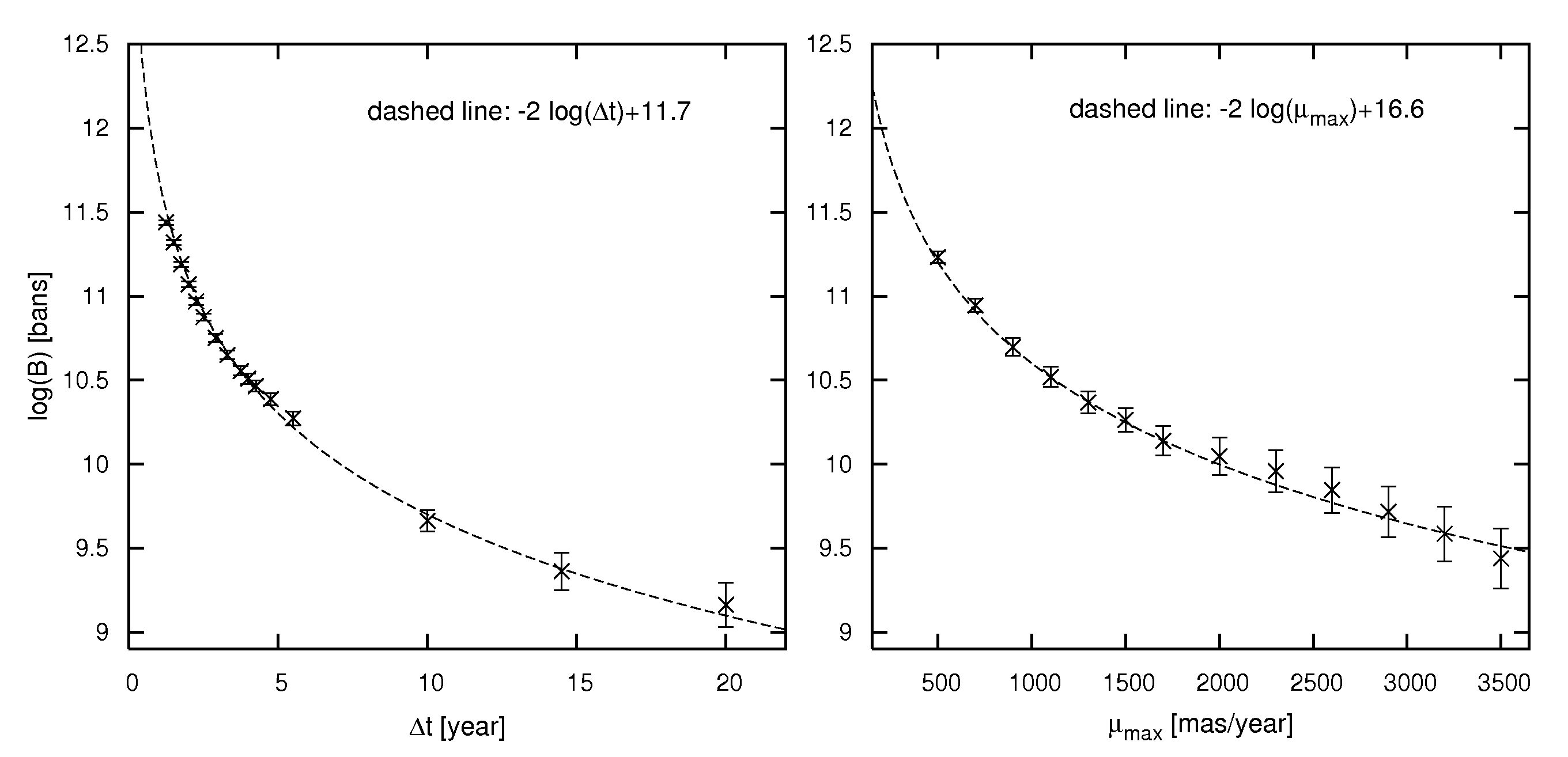}
\caption{Effects of time difference and proper motion limit on Bayes factor in case of uniform prior. 
In the left panel, we consider a mock observation pair with 0.54" separation and vary time intervals. 
The right panel shows the Bayes factor of a 360~mas/yr star with $\Delta{}t=2$ years 
as a function of the proper motion limit; see Equation~(\ref{eq:uconst}). 
The error bars represent the uncertainty of numerical integration.}
\label{uniform}
\end{figure*}

\begin{figure*}
\epsscale{1.1}
\plotone{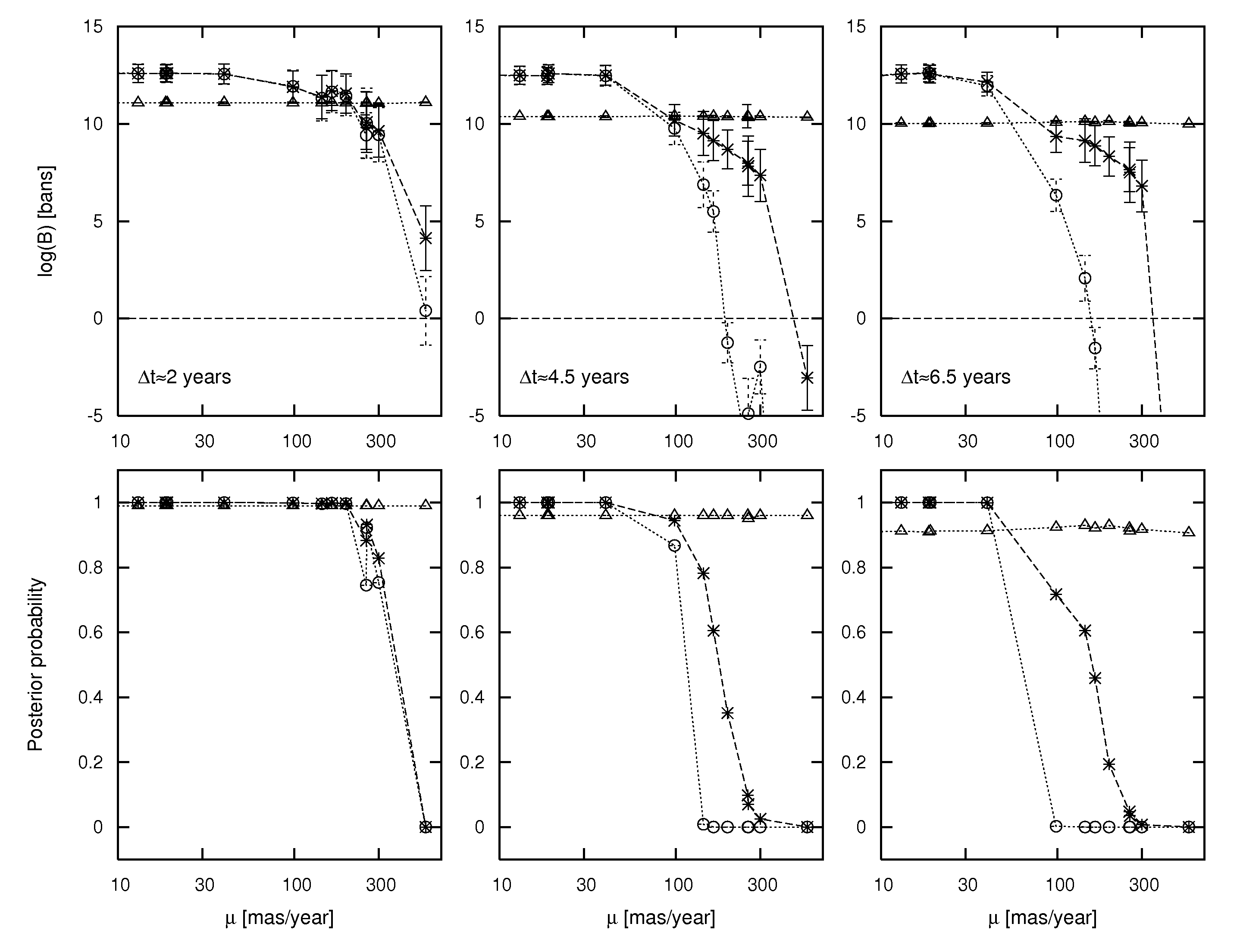}
\caption{Illustration of the weight of evidence (upper panels) and posterior 
probability (lower panels) as a function of proper motion in case of 
2 observations with variable time differences 
(increasing from the left to the right). 
The proper motion is shown on a logarithmic scale. 
Open circles represent the static model, the triangles and crosses correspond to the values from the constant proper motion and the empirical prior, respectively. 
These values are also presented in Table~\ref{tab:dt}.}
\label{2obs}
\end{figure*}

\begin{figure*}
\epsscale{1.1}
\plotone{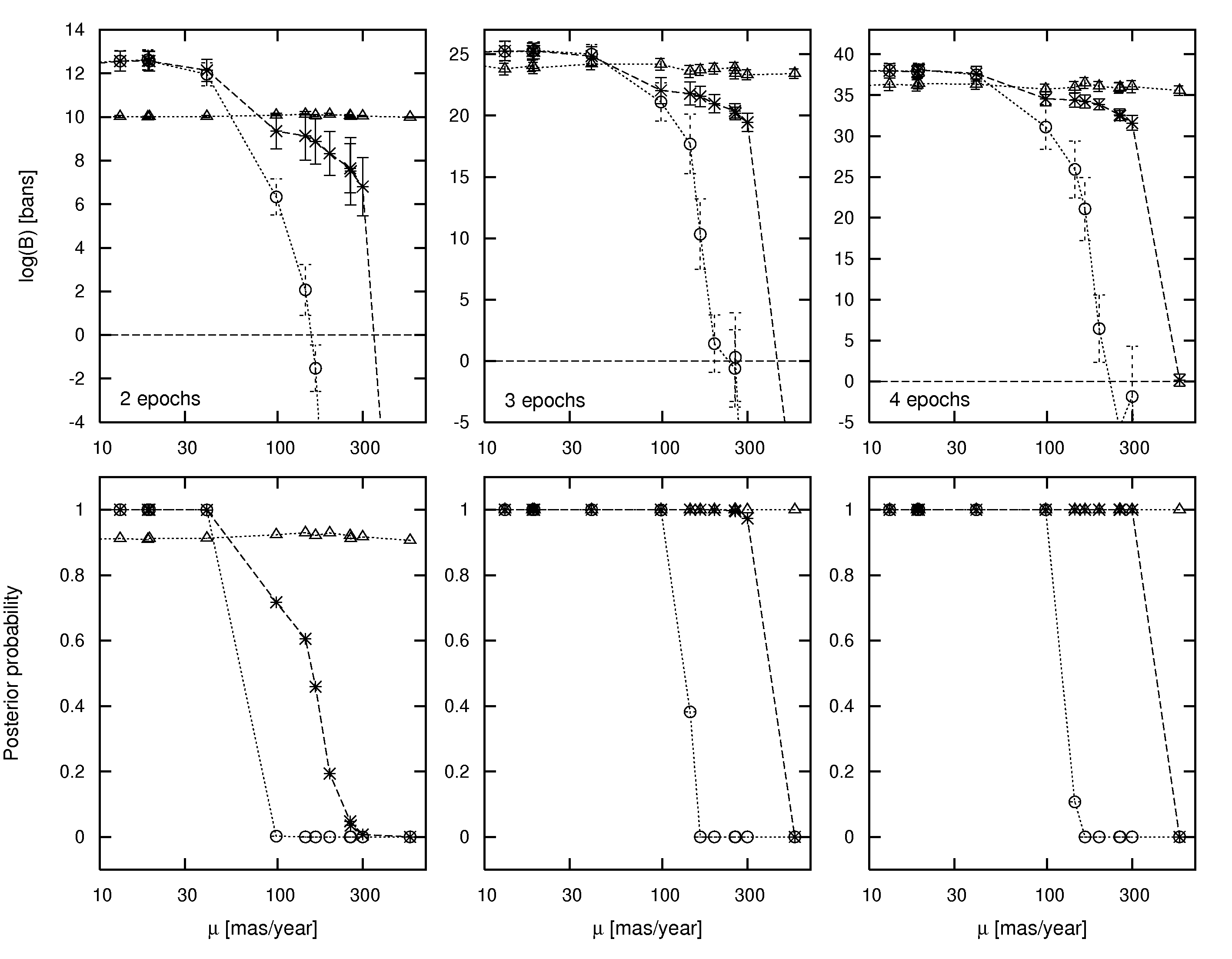}
\caption{The Bayes factor and posterior probability as a function of the 
proper motion in case of 2-, 3- and 4-way associations, 
increasing from left to right, see text.
These values are also presented in Table~\ref{tab:n}.}
\label{long_obs}.
\end{figure*}

\subsection{Ensemble Statistics}

To derive a more realistic $p(\vmu|\vcr,H)$ prior, we choose to study the
ensemble statistics of stars instead of approaching the problem with an
analytic model. While the latter would have the advantage of providing
a function at an arbitrary resolution, the formulas are difficult to derive
and the analytic approximations might miss subtle details of the relation
that could be relevant.

We study the properties of stars in the Sloan Digital Sky Survey
catalog archive that also contains accurate proper motion measurements
from the recalibrated United States Naval Observatory B1.0 Catalog
\citep[USNO-B;][]{Munn2004}.  For this analysis, we pick stars from
the Stripe 82 data set where multiple observations are available over
300 square degrees strip covering a narrow range in declination
between $\pm1.25^{\circ}$.  These repeated observations were taken
between June and December each year from 1998 to 2005 \citep{dr6}.
After rejecting saturated and faint sources 
(u should be in the range of 15--23.5 and g, r, i, z in the range of 14.5--24),
the number of stars is around 100,000.
This size does not allow for a high-resolution determination of the prior, hence we analyze additional
simulation data.

To extend the number of stars used in constructing the prior, we used
the current state-of-the-art Besan\c{c}on models \citep{besancon} that
match the SDSS distributions
well.
Assuming four different stellar populations in the Milky Way, using
the Poisson equation and collisionless Boltzmann equation with a set
of observational parameters (i.e.\ fitting parameters to the dynamical
rotation curve), they compute the number of stars of a given age, type,
effective temperature and absolute magnitude, at any place in the
Galaxy. The model has been successfully used for predictions of
kinematics and comparison with observational data in studies, e.g.,
\citep{bienayme92}, \citep{chareton93}, \citep{ojha99},
\citep{rapaport01}, \citep{soubiran03}.
A total of $740,000$ stars were generated from the Besan\c{c}on models
using large-field equatorial coordinates, thus the prior is dominated
by model data and not by SDSS measurements.
The proper motion distribution of SDSS data and that of the model are
very consistent, with the model yielding somewhat wider distributions
than the observations.

In preparation for binning the data, we separate the dependence of the
prior on the different proper motion components,
$\vmu=(\mu_{\alpha},\mu_{\delta})$, and, omitting the explicit
hypothesis, we write
\begin{equation}\label{PM1}
p(\mu_\alpha,\mu_\delta | \vcr) = p(\mu_\delta|\vcr)\,p(\mu_\alpha|\mu_\delta,\vcr)
\end{equation}
Since the Stripe 82 data set in SDSS contains sources only in
a narrow declination range between $-1.26^{\circ}$ and $+1.26^{\circ}$,
we can safely neglect the dependence on declination in Equation~(\ref{PM1}) for the purpose
of this study, thus
\begin{equation} \label{PM2}
p(\mu_\alpha,\mu_\delta|\vcr) \approx p(\mu_\delta|\alpha)\,p(\mu_\alpha|\mu_\delta,\alpha)
\end{equation}
We establish these relations one-by-one starting with the former using
the basic property of conditional densities
\begin{equation}
p(\mu_\delta|\alpha) = \frac{p(\mu_\delta,\alpha)}{\int\!p(\mu_\delta',\alpha)\,d\mu_\delta'}
\end{equation}
To achieve a better signal-to-noise behaviour across the entire
parameter space, we do not use a uniform grid but vary bin sizes so
that they follow the asinh() in both parameters. In this way one can
have higher resolution bins where more data are available (around the
peak close to 0) and wider bins in the tail.
We further improve the quality of the empirical prior by removing
high-frequency noise with a convolution filter, whose characteristic
width is approximately one pixel in size at any location.
The integral in the denominator is evaluated by counting the stars in the
appropriate bins using the widths of the bins to weight the counts.

Figure~\ref{prior1} shows the prior using the aforementioned
non-linear scale for $\mu_{\delta}$ as function of the position
$\alpha$.
The distribution is centered on approximately
\mbox{$-3\,\rm{mas}/\rm{yr}$}, and the location of the mode is
practically independent of the R.A.
As one nears to the direction of the Galactic Plane
($\alpha=60^{\circ}$ and $\alpha=305^{\circ}$ are the nearest regions
in this stripe) the distribution becomes sharper. This is to be
expected as, if we included a broader range in Right Ascension, the PDF
would get even narrower as the velocity dispersion of stars decreases.

The second term of Equation~(\ref{PM2}) is constructed similarly using
the same adaptive binning and smoothing but in even higher
dimensions. It is difficult to visualize a 4-dimensional PDF, hence,
in Figure~\ref{prior2}, we plot slices of the prior at various
$\alpha$ values.  Both axes are shown in the transformed scale. The
values $\alpha=57.5^{\circ}$ and $\alpha=307.5^{\circ}$ represent the
two edges of Stripe 82, which are closest parts to the Galactic
Plane. The same effect can be seen on these panels as on Figure
\ref{prior1}, looking out of the Plane the PDF gets more disperse. The
boxy (squared) shape of the contour lines arises from the asinh()
transformation; on a linear system, the contours would appear to be
more circular.

\section{Discussion} \label{sec:res}

\subsection{Sample Stars}

As mentioned earlier, Stripe 82 was observed repeatedly from 1998 to
2005, between June and December of each year. Thus we can obtain
multi-epoch observations to test our method. We choose a range of
stars with different proper motions observed at different epochs. To
be sure that for our tests the observed stars are the same in each
epoch we select bright stars (magnitude $r<17$) with tolerances in all
magnitudes (typically 0.7 in u, 0.4 in g, r, i and 0.5 in z). The
query gives us on the average 20 epochs per star, from which many were
observed with small time intervals while the biggest time interval
between the epochs is approximately 6.5 years. We divide the time
interval into 3 approximately equal parts and thus get 4 observations
of each star with much the same time intervals between them. According
to Equations~(\ref{int1}) and (\ref{int2}), only RA and Dec
coordinates are used for calculating the Bayes factors, USNO
measurements of proper motion on the forthcoming figures and tables
are only shown as a reference.
We randomly select a dozen stars for the following tests.

\subsection{Numerical Integration}

We calculate the integrals of the Bayes factor numerically. Our Monte-Carlo implementation
generates independent random positions $\{\vcr_n\}$ (3-D unit vectors) and two random
components of the
velocity that yield the $\vmu_{n}$ vector in the tangent plane of each $\vcr_n$, i.e.,
\begin{equation}
\vcr_n' = \left( \vcr_n + \vmu_n\,\Delta{}t \right) \Big/ \left| \vcr_n + \vmu_n\,\Delta{}t \right|
\end{equation}
In theory, one has to integrate the position over the whole celestial sphere, and the proper motion out to infinity, but the integrands always drop sharply in practice, hence one can
bound all the relevant parameters easily to reduce the computational need and to use the above approximation to the motion. For more efficient implementations, one can utilize more sophisticated Markov chain Monte Carlo (MCMC) methods.

The uncertainty estimates include two separate sources of errors.
The numerical imprecision is tuned by the number of generated random
parameters, and can be estimated in the process of the integration. In
our calculations, this error term is kept at a low level, and contributes $10^{-2}$ order of magnitude to the value of the weight of evidence.

Another source of error comes from the uncertainty in position
measurements. While this is small for the SDSS detections,
$\sigma=0.1"$, the short time differences could yield large relative
errors in the proper motion.  To get the order of this error, we
generate 100 random realizations of the position of every star from
the appropriate Gaussian distribution, and recalculate the Bayes
factor to derive the root-mean-square error in the weight of
evidence. The figures later in this section contain error bars that
represent these 1$\sigma$ deviations.
\newline

\subsection{The Uniform Proper Motion Prior} \label{sec:uniform}

After the static case, the simplest is a uniform prior as introduced
in Equation~(\ref{eq:uconst}).
This analytic formula may appear at first not to favor any particular proper motion,
yet the Bayes factor has some non-trivial scaling properties that
are worth considering.

As the displacement of the source is a product of the time difference and the proper motion,
associations at the same distances but with varying time intervals will indeed have
different Bayes factors.
In the case of a longer $\Delta{}t$, only smaller proper motions will
contribute to the integral in Equation~(\ref{int1}) shrinking the integral domain. This yields a scaling by a factor of $\Delta t^{-2}$,
as seen in the left panel of Figure~\ref{uniform}.
This means that associations will be assigned lower qualities if they
are farther in time even if their angular separations are identical.

Another interesting aspect is the selection of the limiting $\mu_{\rm{max}}$ value.
Our choice of 600~mas/yr is admittedly somewhat arbitrary and was selected
to cover the stars in our sample.
If one decreased its value then stars moving at faster speeds would
quickly get lower Bayes factors and clearly not be associated.
Increasing limits make the value of the constant prior drop,
which in turn will lower the quality of the associations.
For illustrating this effect we compute the Bayes factors for
a star with $\mu=360\ \rm{mas}/\rm{yr}$ and $\Delta{}t=2$ years
as a function of $\mu_{\rm{max}}$, see the right panel of Figure~\ref{uniform}.
As the prior is proportional to $\mu_{\rm{max}}^{-2}$, the curve follows the same trend.


\subsection{The Time Difference}

First we analyze the quality of the associations as a function time difference between the observations.
The top panels of Figure~\ref{2obs} show the logarithm of the Bayes factor, a.k.a.\ the weight of evidence for all stars in our test sample as a function of their proper motion. Open circles represent the results from the static model that can be obtained analytically as in \citet{pxid}, and crosses show the new measurements from the numerical integration of the improved model using the empirical prior introduced in this study. Triangles signal the value for a simple model of constant proper motion prior with $\mu_{\rm{max}}=600\ \rm{mas/yr}$. If we correct for small relative differences in time intervals between the epochs taking 2, 4.5 and 6.5 years as a reference respectively, this prior yields practically constant weights of evidence.
For reference, the $W=0$ threshold is plotted as the dashed horizontal line. This is the theoretical dividing line above which the observations support the hypothesis of the match.
All panels contain the same objects but the calculations are based on different detections that are farther apart in time as we go from left to right.
What we see immediately is that as the time difference increases, the
models provide increasingly different results: the static model
starts rejecting stars with larger proper motions much faster than
models that accommodates the possibility of the sources moving.

While the only objective measure of the quality for the match is the
Bayes factor, its interpretation for the uninitiated is admittedly not
as obvious at first as a probability value would be, where one has a
good sense of the meaning of the values.
From the Bayes factor, we can calculate the posterior $P(H|D)$, if we
have a prior $P(H)$ via the equation
\begin{equation}
P(H|D)=\left[ 1 + \frac{1-P(H)}{B(H,K|D)\,P(H)} \right]^{-1}
\end{equation}
Assuming a constant prior over the sky with the value of $P(H)=1/N$,
where $N=10^9$ is the estimated total number of stars on the sky as
computed from the average density in SDSS, we can plot the matching
probabilities for comparison.
Note that large posterior probabilities are not sensitive to
small modulations in the density;
it changes linearly only for small values of the Bayes factor.
The bottom panels of Figure~\ref{2obs} use the same symbols as the top
ones to illustrate the derived posteriors using the above constant prior.
The difference between the models is possibly even more
striking here: While the left panel has very similar estimates from
the different models, with time the separations grow large enough to
quickly zero out the probabilities for stars with proper motions
larger than 100~mas/yr, whereas the new models keep the
probabilities significantly larger. 
Table~\ref{tab:dt} contains the measurements for all stars as function of the time difference.
The first column of the table is the identifier of the star, the ObjID in SDSS Data Release 6.
The reference proper motion values are taken from \texttt{ProperMotions} table of the SDSS
Catalog Science Archive,
which combine the SDSS and the recalibrated USNO-B astrometry for a
precise and reliable determination.

We see two important features of the associations using the new proper motion
priors. 
The constant prior yields lower and lower Bayes factors
as the elapsed time increases and the probability drops regardless of the
proper motion. Even when the separations are small enough for the static
model to perfectly recover the object, the constant proper motion prior
yields a lower 90\%
probability.
The empirical prior always outperforms the static model but,
in this 2-epoch observation case, the probabilities of the fast
stars fall below the constant case or any reasonable probability threshold.

\subsection{Three and Four Epochs}

Next we turn our attention to the potential improvements from including
additional epochs to the data sets. For this comparison, we keep the first
and last observations in time, hence the baseline is the same for all
cases. We add to these two observations additional measurements whose
epochs are between them in time.
These 2-, 3- and 4-way associations are shown in the left, middle
and right panels of Figure~\ref{long_obs}, respectively.
It is apparent that adding new detections significantly improves the
the proper motion models: the reasonably good associations of only two
detections are promoted to essentially certain matches by including
intermediate detections. In contrast, the static model continues to
reject the associations of all high proper motion stars.
We see that one of the stars with $\mu\sim{}100\ \rm{mas/yr}$
actually gets a high probability even in the static model, when the angular
separation for only a few years between the epochs is small enough to
recover the star.

Table~\ref{tab:n} shows the measurements as a function of the number
of epochs used in the calculations. We see that the empirical prior of
the improved model assigns 100\%
probability of all stars when
considering all 4 epochs, and even the 3-epoch computations would
yield close to that with the $\mu=300\ \rm{mas/yr}$ star getting a lower
97\%.
The exception from this is the fastest star at $\mu=555\ \rm{mas/yr}$
in case of the empirical prior,
whose probability is essentially 0 in all panels. The reason for this
is that this star is one of the highest proper motion stars in Stripe
82 and even with the generated model stars, which appear in the prior,
we have very few (roughly 40) high proper motion stars.

It is worth re-iterating the reason for and the consequence of these results.
Associations of more than two detections
benefit dramatically more from the proper motion prior because two points can always
be connected with a straight line unlike three or more.
In other words, the prior probability of two detections being on a great circle is 100\%
but for three or more it is small, hence such combinations will get boosted
by the alignment.
Having seen the convincingly large probabilities for the 3-way cases
and assuming the same maximum time difference between observations,
one can conclude that, for the time intervals we consider here,
surveying strategies
with 3 epochs are superior to those with only 2, but adding more
would not improve noticeably our ability to correctly cross-identify
the detections.

\section{Conclusions} \label{sec:sum}

We presented an improved model for probabilistic cross-identification
of stars, which accommodates the possibility of moving objects via a
proper motion prior. Using the Bayesian approach of \citet{pxid}, we
performed hypothesis testing with the new models on a sample of SDSS
DR6 stars with known proper motions and compared the results to the
static case.
In accord with our expectations, we found that moving stars would be
missed by association algorithms that neglect to model the motion, but
using an empirical prior of the proper motion would assign larger
observational evidence to the match and higher probabilities. The
dependence of the quality of these cross-identifications was studied
as a function of separation in time (and space) as well as using
multi-epoch observations. The SDSS Stripe 82 sample provided a good
test set with 2--4 detections at different times with a few years in
between. The tests were done assuming a maximum proper motion of 600 mas/yr.
We found that, even though the 2-epoch data sets benefit significantly
from the proper motion model, the 3-epoch observations essentially
recover the right associations even for fast-moving stars, and the
4-epoch cases yield 100\% probabilities. We also conclude, that the empirical prior surpasses the static model for the whole range of proper motions, while the uniform prior performs better only for the high proper-motion stars.
%


Since the analytically computable static case is
still a good model for most celestial sources, it is best
to carry out the cross-identification in multiple steps: first finding
associations using the static model, and then applying the more
computer-intensive proper motion variant only to the remainder of sources.
While it might be tempting to simply increase the positional errors
to discover the associations of moving sources, the procedure would
be far from optimal.
The overall dominant effect of such changes is that the Bayes factor
would drop slower with separation, and,
since the angular distance is essentially divided by the uncertainty,
a ten times larger $\sigma$ would practically yield associations out to ten times
larger distances; most of them incidental.
The improvement of our novel approach over such naive workarounds
comes from using the true uncertainties
and the high sensitivity of the algorithm to sources moving on a
great circle as allowed by the proper motion model.

\acknowledgements %
The authors would like to acknowledge the use of the online tools of
the Besan\c{c}on collaboration to obtain simulated stars for this
study and thank Rosemary Wyse for her invaluable insights and help
with stellar model of the Galaxy.
T.B. acknowledges support from the Gordon and Betty Moore Foundation via GBMF~554. G.K. and I.C. acknowledge support from NKTH:Polanyi, KCKHA005 and OTKA-MB08A-80177. A.C. acknowledges partial support from NSF award AST-0709394.

\def\arraystretch{1.2}

\begin{deluxetable*}{ l c c r r r r r r }
\tablecolumns{7}
\tablewidth{0pt}

\tablecaption{Weights of evidence and posterior probabilities
in the static, uniform prior and the proper-motion models
as a function the elapsed time between two observations \label{tab:dt}}

\tablehead{
\colhead{ObjID}&\colhead{$\mu$}&\colhead{$\Delta t $} &
\multicolumn{3}{c}{Weight} & \multicolumn{3}{c}{Probability}\\
\cline{4-6} \cline{7-9} 
\colhead{} & \colhead{[mas/yr]} & \colhead{[yr]}   & \colhead{static}   & \colhead{uniform}  & \colhead{motion} &
\colhead{static}   & \colhead{uniform}  & \colhead{motion}}
\startdata

587731173305614418 & 13 & 1.38 & 12.59 & 11.08 & 12.59 & 1.00 & 0.99 & 1.00 \\
                     &  & 3.20 & 12.47 & 10.37 & 12.49 & 1.00 & 0.96 & 1.00 \\
                     &  & 4.46 & 12.56 & 10.01 & 12.56 & 1.00 & 0.91 & 1.00 \\  \tableline

587730847429427304 & 18.6 & 2.01 & 12.59 & 11.08 & 12.59 & 1.00 & 0.99 & 1.00 \\
                       &  & 3.95 & 12.48 & 10.37 & 12.47 & 1.00 & 0.96 & 1.00 \\
                       &  & 5.16 & 12.63 & 10.00 & 12.58 & 1.00 & 0.91 & 1.00 \\  \tableline

587731173305876571 & 19 & 1.38 & 12.60 & 11.08 & 12.60 & 1.00 & 0.99 & 1.00 \\
                     &  & 3.20 & 12.58 & 10.37 & 12.58 & 1.00 & 0.96 & 1.00 \\
                     &  & 4.46 & 12.56 & 10.01 & 12.56 & 1.00 & 0.91 & 1.00 \\  \tableline

587731186187763779 & 40 & 2.01 & 12.55 & 11.09 & 12.56 & 1.00 & 0.99 & 1.00 \\
                     &  & 4.03 & 12.46 & 10.36 & 12.49 & 1.00 & 0.96 & 1.00 \\
                     &  & 6.13 & 11.95 & 10.02 & 12.14 & 1.00 & 0.91 & 1.00 \\  \tableline

588015509268725910 & 98 & 2.02 & 11.93 & 11.08 & 11.90 & 1.00 & 0.99 & 1.00 \\
                     &  & 5.15 & 9.76 & 10.41 & 10.18 & 0.87 & 0.96 & 0.94 \\
                     &  & 7.98 & 6.33 & 10.08 & 9.35 & 0.00 & 0.92 & 0.72 \\  \tableline

588015509271805995 & 143 & 2.02 & 11.32 & 11.08 & 11.38 & 1.00 & 0.99 & 1.00 \\
                      &  & 5.07 & 6.88 & 10.38 & 9.50 & 0.01 & 0.96 & 0.78 \\
                      &  & 7.18 & 2.07 & 10.12 & 9.13 & 0.00 & 0.93 & 0.61 \\  \tableline

588015509286813878 & 163 & 2.17 & 11.64 & 11.10 & 11.70 & 1.00 & 0.99 & 1.00 \\
                      &  & 5.01 & 5.50 & 10.38 & 9.13 & 0.00 & 0.96 & 0.61 \\
                      &  & 7.18 & -1.52 & 10.07 & 8.88 & 0.00 & 0.92 & 0.46 \\  \tableline

588015509268201645 & 196 & 2.02 & 11.44 & 11.07 & 11.56 & 1.00 & 0.99 & 1.00 \\
                      &  & 6.11 & -1.24 & 10.36 & 8.68 & 0.00 & 0.96 & 0.35 \\
                      &  & 7.98 & -11.97 & 10.12 & 8.33 & 0.00 & 0.93 & 0.19 \\  \tableline

588015509273378938 & 255 & 2.04 & 9.41 & 11.05 & 9.82 & 0.75 & 0.99 & 0.88 \\
                      &  & 5.08 & -7.07 & 10.39 & 7.98 & 0.00 & 0.96 & 0.10 \\
                      &  & 7.20 & -24.16 & 10.06 & 7.65 & 0.00 & 0.92 & 0.05 \\  \tableline

588015509279342731 & 257 & 2.02 & 10.02 & 11.08 & 10.08 & 0.92 & 0.99 & 0.93 \\
                      &  & 5.17 & -4.89 & 10.32 & 7.82 & 0.00 & 0.95 & 0.07 \\
                      &  & 7.18 & -24.11 & 10.01 & 7.51 & 0.00 & 0.91 & 0.04 \\  \tableline

587730847426740272 & 300 & 1.94 & 9.43 & 11.04 & 9.63 & 0.75 & 0.99 & 0.83 \\
                      &  & 4.01 & -2.49 & 10.37 & 7.35 & 0.00 & 0.96 & 0.02 \\
                      &  & 6.13 & -22.84 & 10.04 & 6.81 & 0.00 & 0.91 & 0.01 \\  \tableline

588015509283930154 & 555 & 2.19 & 0.40 & 11.09 & 4.12 & 0.00 & 0.99 & 0.00 \\
                      &  & 4.11 & -36.83 & 10.33 & -3.05 & 0.00 & 0.96 & 0.00 \\
                      &  & 7.18 & -139.78 & 9.99 & -21.80 & 0.00 & 0.91 & 0.00  
\enddata
\end{deluxetable*}

\def\arraystretch{1.2}
\begin{deluxetable*}{ l c c r r r r r r }
\tablecolumns{7}
\tablewidth{0pt}

\tablecaption{Weights of evidence and posterior probabilities
in the static, uniform prior and proper-motion models for the 2-, 3- and 4-way associations \label{tab:n}}

\tablehead{
\colhead{ObjID} &   \colhead{$\mu$}   & \colhead{$N_{\rm{obs}}$} &
\multicolumn{3}{c}{Weight} & \multicolumn{3}{c}{Probability}\\
\cline{4-6} \cline{7-9} 
\colhead{} & \colhead{[mas/yr]} & \colhead{}   & \colhead{static}  & \colhead{uniform}  & \colhead{motion} &
 \colhead{static}  & \colhead{uniform}  & \colhead{motion}}

\startdata

587731173305614418 & 13 & 2 & 12.56 & 10.01 & 12.56 & 1.00 & 0.91 & 1.00 \\
					 &  & 3 & 25.25 & 23.80 & 25.28 & 1.00 & 1.00 & 1.00 \\
					 &  & 4 & 37.99 & 36.33 & 37.98 & 1.00 & 1.00 & 1.00 \\  \tableline

587730847429427304 & 18.6 & 2 & 12.63 & 10.00 & 12.58 & 1.00 & 0.91 & 1.00 \\
					 &  & 3 & 25.24 & 24.06 & 25.27 & 1.00 & 1.00 & 1.00 \\
					 &  & 4 & 37.91 & 36.19 & 37.86 & 1.00 & 1.00 & 1.00 \\  \tableline

587731173305876571 & 19 & 2 & 12.56 & 10.01 & 12.56 & 1.00 & 0.91 & 1.00 \\
					 &  & 3 & 25.32 & 23.89 & 25.27 & 1.00 & 1.00 & 1.00 \\
					 &  & 4 & 38.11 & 36.41 & 38.08 & 1.00 & 1.00 & 1.00 \\  \tableline

587731186187763779 & 40 & 2 & 11.95 & 10.02 & 12.14 & 1.00 & 0.91 & 1.00 \\
					 &  & 3 & 25.01 & 24.21 & 24.84 & 1.00 & 1.00 & 1.00 \\
					 &  & 4 & 37.48 & 36.34 & 37.64 & 1.00 & 1.00 & 1.00 \\  \tableline

588015509268725910 & 98 & 2 & 6.33 & 10.08 & 9.35 & 0.00 & 0.92 & 0.72 \\
					 &  & 3 & 21.11 & 24.18 & 22.05 & 1.00 & 1.00 & 1.00 \\
					 &  & 4 & 31.10 & 35.74 & 34.56 & 1.00 & 1.00 & 1.00 \\  \tableline

588015509271805995 & 143 & 2 & 2.07 & 10.12 & 9.13 & 0.00 & 0.93 & 0.61 \\
					 &  & 3 & 17.69 & 23.60 & 21.79 & 0.38 & 1.00 & 1.00 \\
					 &  & 4 & 25.92 & 35.96 & 34.41 & 0.11 & 1.00 & 1.00 \\  \tableline

588015509286813878 & 163 & 2 & -1.52 & 10.07 & 8.88 & 0.00 & 0.92 & 0.46 \\
					 &  & 3 & 10.33 & 23.71 & 21.56 & 0.00 & 1.00 & 1.00 \\
					 &  & 4 & 21.10 & 36.48 & 34.22 & 0.00 & 1.00 & 1.00 \\  \tableline

588015509268201645 & 196 & 2 & -11.97 & 10.12 & 8.33 & 0.00 & 0.93 & 0.19 \\
					 &  & 3 & 1.41 & 23.82 & 20.97 & 0.00 & 1.00 & 1.00 \\
					 &  & 4 & 6.46 & 36.08 & 33.88 & 0.00 & 1.00 & 1.00 \\  \tableline

588015509273378938 & 255 & 2 & -24.16 & 10.06 & 7.65 & 0.00 & 0.92 & 0.05 \\
					 &  & 3 & -0.61 & 23.88 & 20.39 & 0.00 & 1.00 & 1.00 \\
					 &  & 4 & -5.76 & 35.87 & 32.60 & 0.00 & 1.00 & 1.00 \\  \tableline

588015509279342731 & 257 & 2 & -24.11 & 10.01 & 7.51 & 0.00 & 0.91 & 0.04 \\
					 &  & 3 & 0.30 & 23.44 & 20.18 & 0.00 & 1.00 & 0.99 \\
					 &  & 4 & -5.44 & 35.95 & 32.49 & 0.00 & 1.00 & 1.00 \\  \tableline

587730847426740272 & 300 & 2 & -22.84 & 10.04 & 6.81 & 0.00 & 0.92 & 0.01 \\
					 &  & 3 & -21.3 & 23.32 & 19.46 & 0.00 & 1.00 & 0.97 \\
					 &  & 4 & -1.85 & 36.05 & 31.60 & 0.00 & 1.00 & 1.00 \\  \tableline

588015509283930154 & 555 & 2 & -139.78 & 9.99 & -21.80 & 0.00 & 0.91 & 0.00 \\
					 &  & 3 & -37.41 & 23.42 & -11.46 & 0.00 & 1.00 & 0.00 \\
					 &  & 4 & -129.09 & 35.56 & 0.17 & 0.00 & 1.00 & 0.00   

\enddata
\end{deluxetable*}

\end{document}